\documentclass{ws-mpla}

\usepackage{amsmath}
\usepackage{amssymb}
\usepackage{bm}
\usepackage{graphicx}
\usepackage{MnSymbol}
\usepackage{hyperref}

\newcommand{\A}{{\bf A}}
\newcommand{\B}{{\bf B}}

\newcommand{\E}{{\bf E}}

\newcommand{\n}{{\bf \nabla}}

\newcommand{\pp}{{\bf p}}

\newcommand{\rr}{{\bf r}}
\newcommand{\tF}{\tilde{F}}
\newcommand{\ii}{\hat{\bf i}}
\newcommand{\jj}{\hat{\bf j}}
\newcommand{\kk}{\hat{\bf k}}
\newcommand{\EE}{\hat{\bf E}}
\newcommand{\BB}{\hat{\bf B}}
\newcommand{\Ah}{\hat{\bf A}}
\newcommand{\ph}{\hat{\phi}}

\begin{document}

\markboth{LUCA VISINELLI}
{AXION-ELECTROMAGNETIC WAVES}

\catchline{}{}{}{}{}

\title{AXION-ELECTROMAGNETIC WAVES}

\author{\footnotesize LUCA VISINELLI}

\address{Department of Physics and Astronomy,\\ The University of Utah, 115 South 1400 East \#201,\\ Salt Lake City, Utah 84112-0830, USA\\
luca.visinelli@utah.edu}

\maketitle

\pub{Received: \today}{}

\begin{abstract}
We extend the duality symmetry between the electric and the magnetic fields to the case in which an additional axion-like term is present, and we derive the set of Maxwell's equations that preserves this symmetry. This new set of equations allows for a gauge symmetry extending the ordinary symmetry in the classical electrodynamics. We obtain explicit solutions for the new set of equations in the absence of external sources, and we discuss the implications of a new internal symmetry between the axion field and the electromagnetic gauge potential.
\end{abstract}

\ccode{PACS Nos.: 14.80.Va, 03.50.De}

\section{Introduction}	

It has long been known that source-free Maxwell's equations for an electro-magnetic (EM) field show an internal symmetry, known as the duality transformation. In a nutshell, Maxwell's equations are invariant when the electric and the magnetic fields $\E$ and $\B$ are mixed by a rotation of an arbitrary angle $\xi$, as
\begin{equation} \label{rotation_EM}
\left( \begin{array}{c}
 {\bf E'}\\
 {\bf B'}\\
 \end{array} \right)  =
\left( \begin{array}{cc}
 \cos\xi & \sin\xi\\
 -\sin\xi & \cos\xi\\
 \end{array} \right)\,\left( \begin{array}{c}
 \E\\
 \B\\
 \end{array} \right).
\end{equation}
This symmetry can be extended to Maxwell's equations in the presence of sources, provided that additional magnetic charges and currents are introduced in the theory. The duality symmetry has been proven to be associated with a set of conserved currents, as first discussed by Lipkin \cite{lipkin} (see also Refs.~\cite{morgan} \cite{kibble} \cite{fradkin} \cite{karch}).

Whenever a pseudoscalar axion-like field $\theta = \theta(x)$ is introduced in the theory, the dual symmetry is spontaneously and explicitly broken. Axion-like fields and their interactions with the EM field have been intensively studied~\cite{anderson} \cite{rosen} \cite{sikivie} \cite{huang} \cite{wilczek} \cite{gasperini} \cite{jacobs}, and they have recently received attention due to the possible role they might play in building topological insulators. However, the set of equations used in the axion electrodynamics literature does not preserve the duality relation in Eq.~(\ref{rotation_EM}). This fact has not been previously noticed, and it might be due to the fact that the introduction of an axion-like interaction with the electromagnetic field only modifies two of the four Maxwell's equations, namely Gauss and Ampere laws; the remaining two equations (Faraday law and Gauss law for $\B$) are usually coupled to the new Gauss and Ampere laws without modifications from the original Maxwell set.

%
In this paper, we derive Maxwell's equations in the presence of an axion-like field, with the requirement that the electric and magnetic fields satisfy the duality relation in Eq.~(\ref{rotation_EM}). Along with Gauss and Ampere laws for axion electrodynamics, we obtain new terms that also modify Faraday law and Gauss law for $\B$. We show that, when the electric charges and currents are set to zero, Maxwell's equations for axion electrodynamics allow for a solution in which coupled axion and electromagnetic waves propagate.

In order to fix our notation, we first review classical electromagnetism in the presence of an electric charge distribution $\rho_e$ and an electric current density ${\bf J}_e$. This paper is organized as follows. In Sec.~\ref{Review of axion electrodynamics}, we review Maxwell's formulation of classical EM, the duality symmetry within classical EM, and the inclusion of a pseudo-scalar axion-like field to the Maxwell's Lagrangian for electromagnetism. In Sec.~\ref{Imposing the duality symmetry}, we obtain a new set of equations describing the EM field configuration when both electric and magnetic sources, together with an axion-like field, are considered. This set of equations is derived by imposing that the equations for the axion electrodynamics be invariant under the duality transformation in Eq.~(\ref{rotation_EM}). Sec.~\ref{Propagation of waves in the axion electrodynamics} is devoted to expressing a solution in free space for the axion electrodynamics. We obtain one explicit solution in which the axion field is described by a Klein-Gordon equation of motion, with the electric and magnetic fields propagating orthogonally and coupling to the axion field. In Sec.~\ref{Gauge transformations in the axion electrodynamics}, we derive the expression for the gauge potentials in the theory, and we express the equations for the axion electrodynamics in terms of these potentials. Finally, in Sec.~(\ref{Conservation of currents}), we show that an internal symmetry between the axion field and the gauge potential exists, and we derive a set of conserved current related to this symmetry.

\section{Review of axion electrodynamics} \label{Review of axion electrodynamics}

\subsection{Classical electrodynamics}

In the vacuum, Maxwell's equations for an EM field with electric field $\E$, magnetic field $\B$, and the sources $\rho_e$, ${\bf J_e}$ are
\begin{equation}\label{maxwell_equations}
\begin{array}{l}
\n \cdot \E = \rho_e/\epsilon_0,\\
\n \times \B -  \epsilon_0\,\mu_0\,\partial_t\,\E = \mu_0\,{\bf J_e},\\
\n \cdot \B = 0,\\
\n \times \E + \partial_t\,\B = 0,
\end{array}
\end{equation}
where $\epsilon_0$ and $\mu_0$ are the electric polarizability and the magnetic permeability of the vacuum, respectively. The speed of light in the vacuum is $c = 1/\sqrt{\epsilon_0\,\mu_0}$. Thanks to the third and the fourth line in Eq.~(\ref{maxwell_equations}), it is possible to write the electric and the magnetic fields in terms of a scalar potential $\phi$ and a vector potential ${\bf A}$, as
\begin{equation} \label{gauge_classic}
\E = -\partial_t\,\bf A - \n\phi,\quad\hbox{and}\quad \B = \n \times {\bf A}.
\end{equation}

Maxwell's Eqs.~(\ref{maxwell_equations}) can be reformulated in an elegant form that is explicitly covariant. Firstly, we notice that the source $J_e^\mu = (c\,\rho_e, {\bf J_e})$ and the potential $A^{\mu} = (\phi/c, {\bf A})$ transform as four-vectors under a Lorentz transformation. Once the antisymmetric field tensor and its dual are being introduced as
\begin{equation}\label{def_F}
F^{\mu\nu} = \partial^\mu A^\nu-\partial^\nu A^\mu, \quad\hbox{and}\quad \tF^{\mu\nu} = \frac{1}{2}\,\epsilon^{\mu\nu\sigma\rho}\,F_{\sigma\rho},
\end{equation}
the electric and magnetic fields read
\begin{equation} \label{relation_strength_fields}
E^i = c\,F^{i0} = -\frac{d\phi}{dx^i}-\frac{dA^i}{dt},\quad\hbox{and}\quad B^i = \frac{1}{2}\,\epsilon^{ijk}\,F_{jk} = (\n\times {\bf A})^i.
\end{equation}
With this notation, Maxwell's equations are rewritten as
\begin{equation} \label{covariant_eq}
\begin{array}{l}
\partial_\mu\,F^{\mu\nu} = \mu_0\,J_e^\nu,\\
\partial_\mu\,\tF^{\mu\nu} = 0.
\end{array}
\end{equation}
The first line of Maxwell's Eqs.~(\ref{covariant_eq}) can be obtained from Euler's equation of motion,
\begin{equation}\label{variational_euler}
\partial^\mu\,\frac{\partial \mathcal{L}_0}{\partial \,(\partial^\mu\,A^\nu)} = \frac{\partial \mathcal{L}_0}{\partial A^\nu},
\end{equation}
using the EM Lagrangian $\mathcal{L}_0 = -(1/4\,\mu_0)\,F^{\mu\nu}\,F_{\mu\nu} - A_\mu\,J_e^\mu$.
However, the second line in Eq.~(\ref{covariant_eq}) does not directly derive from the Lagrangian principle, being motivated empirically by the non-discovery of magnetic monopoles ($\n \cdot \B = 0$), and the absence of magnetic currents. 

\subsection{Duality symmetry and magnetic sources} \label{Duality symmetry and magnetic sources}

When a magnetic charge density $\rho_m$ and a magnetic current ${\bf J_m}$ are included in the theory, Maxwell's Eqs.~(\ref{maxwell_equations}) appear in the symmetric form
\begin{equation}\label{maxwell_eq_monopole}
\begin{array}{l}
\n \cdot \E = \rho_e/\epsilon_0,\\
\n \times \B -  \epsilon_0\,\mu_0\,\partial_t\,\E = \mu_0\,{\bf J_e},\\
\n \cdot \B = \mu_0\,\rho_m,\\
\n \times \E + \partial_t\,\B = -\mu_0\,{\bf J_m}.
\end{array}
\end{equation}
Similarly to what shown in Eq.~(\ref{covariant_eq}), this last set of equations can be expressed in a covariant form. In fact, defining $J_m^\mu = (c\,\rho_m, {\bf J_m})$, we have
\begin{equation} \label{covariant_eq_monopole}
\begin{array}{l}
\partial_\mu\,F^{\mu\nu} = \mu_0\,J_e^\nu,\\
\partial_\mu\,\tF^{\mu\nu} = \mu_0\,J_m^\nu/c.
\end{array}
\end{equation}
Taking the derivative of each line in Eq.~(\ref{covariant_eq_monopole}) in $x^\nu$ yields to the conservation of the electric and magnetic currents,
\begin{equation} \label{conservation_current}
\partial_\nu\,J_e^\nu = 0,\quad \hbox{and}\quad \partial_\nu\,J_m^\nu = 0.
\end{equation}
This result is obtained by noticing that, since $\partial_\mu\partial_\nu$ is a symmetric tensor operator, we have $\partial_\mu\partial_\nu\,F^{\mu\nu} = \partial_\mu\partial_\nu\, \tF^{\mu\nu} = 0$. Introducing the SO(2) matrix
\begin{equation}
U(\xi) = \left( \begin{array}{cc}
 \cos\xi & \sin\xi\\
 -\sin\xi & \cos\xi\\
 \end{array} \right),
\end{equation}
the field strength and its dual in Eq.~(\ref{covariant_eq_monopole}) are invariant under a rotation by an angle $\xi$,
\begin{equation} \label{rotation_fields}
\left( \begin{array}{c}
 F^{'\mu\nu}\\
 \tF^{'\mu\nu}\\
 \end{array} \right)  =
U(\xi)\,\left( \begin{array}{c}
 F^{\mu\nu}\\
 \tF^{\mu\nu}\\
 \end{array} \right),
\end{equation}
when the sources are rotated by
\begin{equation}  \label{rotation_sources}
\left( \begin{array}{c}
 J_e^{'\mu}\\
 J_m^{'\mu}/c\\
 \end{array} \right)  =
U(\xi)\,\left( \begin{array}{c}
 J_e^{\mu}\\
 J_m^{\mu}/c\\
  \end{array} \right).
\end{equation}
Notice that the transformations in Eqs.~(\ref{rotation_EM}) and~(\ref{rotation_fields}) are equivalent. Since magnetic monopoles and currents are not observed in nature, we choose the angle $\bar{\xi}$ for which, given a certain set of currents $J_e^\mu$ and $J_m^\mu$, makes $J_m^{'\mu}$ disappear in the rotated configuration. Using Eq.~(\ref{rotation_sources}), this angle is defined by
\begin{equation}
J_e^\mu\,\sin\bar{\xi} = J_m^\mu\,\cos\bar{\xi}.
\end{equation}
To sum up, although the theory is invariant under the transformations in Eqs.~(\ref{rotation_fields}) and~(\ref{rotation_sources}), empirical results force the physical choice $J_m^\mu = 0$. Nevertheless, we wish to keep the SO(2) symmetry even when an axion-like term is added to the electromagnetic Lagrangian.

\subsection{Adding an axion-like term} \label{Adding an axion-like term}

Maxwell's Eqs.~(\ref{maxwell_equations}) modify if the additional axion-like term
\begin{equation}\label{axion_Lagrangian}
\mathcal{L}_\theta = -\frac{\kappa}{\mu_0\,c}\,\theta\,\E\cdot\B,
\end{equation}
is added to the EM Lagrangian $\mathcal{L}_0$. In Eq.~(\ref{axion_Lagrangian}), $\theta = \theta(x)$ is a pseudo-scalar field known in the particle physics literature as the axion-like field, and $\kappa$ is a coupling constant; the factor $1/\mu_0\,c$ assures that $\kappa$ is dimensionless. The effects to Maxwell's equations, of the axion-like term above, have been long discussed~\cite{sikivie} \cite{huang} \cite{wilczek} \cite{gasperini} \cite{jacobs}. Notice that the appearance of the axion-like term in Eq.~(\ref{axion_Lagrangian}) explicitly breaks the SO(2) symmetry in Eq.~(\ref{rotation_EM}), unless we are at the CP-preserving configuration $\bar{\theta} = 0$.

When we consider the EM Lagrangian with the term in Eq.~(\ref{axion_Lagrangian}) added, the resulting equations of motion for the $\E$ and $\B$ fields are Maxwell's Eqs.~(\ref{maxwell_equations}), with the electric charge and current densities replaced by
\begin{equation} \label{substitution}
\rho_e \to \rho_e + \frac{\kappa}{\mu_0\,c}\, \B \cdot\n\theta, \quad \hbox{and}\quad {\bf J_e} \to {\bf J_e} -\frac{\kappa}{\mu_0\,c}\,\left[(\partial_t\,\theta)\,\B + \n\theta\times\E\right].
\end{equation}
The axion-like term in Eq.~(\ref{axion_Lagrangian}) can be written in a manifestly covariant form as 
\begin{equation}\label{axion_Lagrangian1}
\mathcal{L}_\theta =  -\frac{\kappa}{\mu_0\,c}\,\theta\,\E\cdot\B = \frac{\kappa}{4\mu_0}\,\theta\,F_{\mu\nu}\,\tF^{\mu\nu},
\end{equation}
and the Lagrangian for the axion electrodynamics is
\begin{equation}\label{Lagrangian_axion}
\mathcal{L}_{0+\theta} = -\frac{1}{4\,\mu_0}\,F^{\mu\nu}\,F_{\mu\nu} + \frac{\kappa}{4\mu_0}\,\theta\,F_{\mu\nu}\,\tF^{\mu\nu} - A_\mu\,J_e^\mu + \mathcal{L}_a,
\end{equation}
where the axion Lagrangian with a potential $U(\theta)$ is
\begin{equation}
\mathcal{L}_a = \frac{1}{2}\,\left(\partial^\mu\,\theta\right)\,\left(\partial_\mu\,\theta\right) - U(\theta).
\end{equation}
Using the Euler-Lagrange equations for $\mathcal{L}_{0 + \theta}$,
\begin{equation} \label{term1.1}
\partial_\mu\,\frac{\partial\mathcal{L}_{0 +\theta}}{\partial\,(\partial_\mu A_\nu)} = \frac{\partial\mathcal{L}_{0+\theta}}{\partial\,A_\nu},\quad \hbox{and}\quad
\partial_\mu\,\frac{\partial\mathcal{L}_{0+\theta}}{\partial\,(\partial_\mu \theta)} = \frac{\partial\mathcal{L}_{0+\theta}}{\partial\,\theta},
\end{equation}
we obtain the dynamical equations of motion for the electromagnetic field
\begin{equation} \label{axion_field_equations}
\begin{array}{l}
\partial_\mu \, F^{\mu\nu} - \kappa\,\partial_\mu\left(\theta\,\tF^{\mu\nu}\,\right) = \mu_0\,J_e^\nu,\\
\Box\,\theta = \frac{\kappa}{4\mu_0}\,F_{\mu\nu}\,\tF^{\mu\nu} - \frac{\partial U(\theta)}{\partial\theta},
\end{array}
\end{equation}
where $\Box\theta = \partial_{tt}\,\theta - \n^2\,\theta$.

\section{Equations for the axion electrodynamics} \label{New equations for the axion electrodynamics}

\subsection{Imposing the duality symmetry} \label{Imposing the duality symmetry}

We now derive the set of Maxwell's equations that takes into account the axion-like term introduced in Sec.~\ref{Adding an axion-like term}, and is invariant under the duality transformation discussed in Sec.~\ref{Duality symmetry and magnetic sources}. In fact, the Lagrangian for the axion electrodynamics in Eq.~(\ref{Lagrangian_axion}) allows us to derive the set of Eqs.~(\ref{axion_field_equations}), but does not provide any information on the additional equation that the dual tensor $F^{\mu\nu}$ satisfies. Usually, in the axion electrodynamics literature, it is assumed the constraint
\begin{equation} \label{additional_constraint}
\partial_\mu \,\tF^{\mu\nu} = 0,
\end{equation}
which is the same as the one in Maxwell's Eq.~(\ref{covariant_eq}) in the absence of an axion term and magnetic sources. However, this choice makes the set of Eqs.~(\ref{axion_field_equations}) and~(\ref{additional_constraint}) not invariant under the SO(2) duality symmetry.

Here, we impose the duality symmetry to the  in the axion-like theory as well, and we replace the condition in Eq.~(\ref{additional_constraint}) with a dynamical equation for $\tilde{F}^{\mu\nu}$. We start by noticing that, in the symmetric Eq.~(\ref{covariant_eq_monopole}), the second line is obtained from the first line by using the transformations in Eq.~(\ref{rotation_fields}) and~(\ref{rotation_sources}) with $\xi = \pi/2$. Applying the same transformations to the first line of Eq.~(\ref{axion_field_equations}), we obtain
\begin{equation} \label{additional_field_eq}
\partial_\mu\,\tF^{\mu\nu} + \kappa\,\partial_\mu\left(\theta\,F^{\mu\nu}\,\right) = \mu_0\,J_m^\nu/c.
\end{equation}
Together, Eqs.~(\ref{axion_field_equations}) and~(\ref{additional_field_eq}) relate the field strengths $F^{\mu\nu}$ and $\tF^{\mu\nu}$ to the electric and magnetic currents, $J_e^\mu$ and $J_m^\mu$, and to the axion-like field $\theta$. Summing up, the complete set of equations for the axion-electromagnetic field is
\begin{equation} \label{maxwell_axion_complete}
\begin{array}{l}
\partial_\mu\,F^{\mu\nu} - \kappa\,\partial_\mu\left(\theta\,\tF^{\mu\nu}\,\right) = \mu_0\,J_e^\nu,\\
\partial_\mu\,\tF^{\mu\nu} + \kappa\,\partial_\mu\left(\theta\,F^{\mu\nu}\,\right) = \mu_0\,J_m^\nu/c,\\
\Box\,\theta = \frac{\kappa}{4\mu_0}\,F_{\mu\nu}\,\tF^{\mu\nu} - \frac{\partial U(\theta)}{\partial\theta}.
\end{array}
\end{equation}
In terms of the electric and magnetic fields $\E$ and $\B$, Eq.~(\ref{maxwell_axion_complete}) rewrites as
\begin{equation} \label{maxwell_axion_complete_1}
\begin{array}{l}
\n \cdot \left(\E - c \kappa \, \theta\,\B\right) = \rho_e/\epsilon_0,\\
\n \times \left( c\B + \kappa \, \theta \,\E \right) = \partial_t\left( \E - c \kappa \, \theta\,\B\right)/c + c \mu_0\,{\bf J}_e,\\
\n \cdot \left( c\B + \kappa \, \theta\,\E \right) = c\mu_0 \,\rho_m,\\
\n \times \left(\E - c \kappa \, \theta\,\B\right) = -\partial_t\left( c\B + \kappa \, \theta\,\E\right)/c -\mu_0\,{\bf J}_m,\\
\Box\,\theta = -\frac{\kappa}{\mu_0\,c}\,\E\cdot\B - \frac{\partial U(\theta)}{\partial\theta}.
\end{array}
\end{equation}
Rotating the fields and sources by an angle $\bar{\xi}$, so that $\rho_m$ and ${\bf J}_m$ vanishes, we finally obtain Maxwell's equations in the presence of an axion-like field
\begin{equation} \label{maxwell_axion1}
\begin{array}{l}
\n \cdot \left(\E - c \kappa \, \theta\,\B\right) = \rho_e/\epsilon_0,\\
\n \times \left( c\B + \kappa \, \theta \,\E \right) = \partial_t\left( \E - c \kappa \, \theta\,\B\right)/c + c \mu_0\,{\bf J}_e,\\
\n \cdot \left( c\B + \kappa \, \theta\,\E \right) = 0,\\
\n \times \left(\E - c \kappa \, \theta\,\B\right) + \partial_t\left( c\B + \kappa \, \theta\,\E\right)/c = 0,\\
\Box\,\theta = -\frac{\kappa}{\mu_0\,c}\,\E\cdot\B - \frac{\partial U(\theta)}{\partial\theta}.
\end{array}
\end{equation}
This set of equations for the axion electrodynamics was first derived in Refs.~\cite{sudbery} \cite{tiwari} using a four-vector Lagrangian. Instead, here we have avoided complications due to a vector Lagrangian, and we have derived this set of equations by using the gauge symmetry in Eq.~(\ref{rotation_EM}).

\subsection{Propagation of waves in the axion electrodynamics}\label{Propagation of waves in the axion electrodynamics}

From here to the end of the paper, we switch to natural units. In order to consider the propagation of axion-electromagnetic waves in free space, we fix the external sources $\rho_e = {\bf J}_e = 0$. Assuming an axion potential of the form
\begin{equation} \label{quadratic_potential}
U(\theta) = \frac{1}{2}\,m^2\,\theta^2,
\end{equation}
with $m$ the axion mass, the set of wave Eq.~(\ref{maxwell_axion1}) in free space reads
\begin{equation} \label{maxwell_axion_free_wave}
\begin{array}{l}
\n \cdot \left(\E - \kappa\, \theta\, \B\right) = 0,\\
\n \cdot \left(\B + \kappa\, \theta\, \E\right) = 0,\\
\n \times \left(\E - \kappa\, \theta\, \B\right) + \partial_t \,\left(\B + \kappa\, \theta\, \E\right) = 0,\\
\n \times \left(\B + \kappa\, \theta\, \E\right) - \partial_t \,\left(\E - \kappa\, \theta\, \B\right) = 0,\\
\left(\Box + m^2\right) \,\theta = -\kappa\,\E\cdot\B.
\end{array}
\end{equation}
Taking the curl of the third and fourth lines in Eq.~(\ref{maxwell_axion_free_wave}) and rearranging, we obtain
\begin{equation} \label{maxwell_axion_free_wave1}
\begin{array}{l}
\n \cdot \,\left(\E - \kappa\, \theta\, \B\right) = 0,\\
\n \cdot \,\left(\B + \kappa\, \theta\, \E\right) = 0,\\
\Box \, \left(\E - \kappa\, \theta\, \B\right) = 0,\\
\Box \,\left(\B + \kappa\, \theta\, \E\right) = 0,\\
\left(\Box + m^2\right) \theta = -\kappa\,\E\cdot\B.
\end{array}
\end{equation}
Setting
\begin{equation} \label{definition_hatEB}
\begin{array}{l}
\EE = \E - \kappa\, \theta\, \B,\\
\BB = \B + \kappa\, \theta\, \E,
\end{array}
\quad\quad\hbox{satisfying}\quad\quad
\begin{array}{l}
\Box\,\EE = 0,\\
\Box\BB = 0,
\end{array}
\end{equation}
one solution to the set of wave equations for the fields $\EE$ and $\BB$ is
\begin{equation} \label{solution_hatEB}
\begin{array}{l}
\EE = I_+\,\cos(kz - \omega t)\,\ii - I_-\,\sin(kz - \omega t)\, \jj,\\
\BB = I_-\,\sin(kz - \omega t)\,\ii + I_+\,\cos(kz - \omega t)\,\jj.
\end{array}
\end{equation}
This solution describes transverse plane waves propagating along the $z$-direction with momentum $k$, frequency $\omega$, and amplitude governed by $I_+$ and $I_-$. The wave equations for $\EE$ and $\BB$ force $\omega = k$. The corresponding values of the fields $\E$ and $\B$ are found by inverting Eq.~(\ref{definition_hatEB}), obtaining
\begin{equation} \label{solution_EB}
\E = \frac{\EE + \kappa\,\theta\,\BB}{1+\kappa^2\,\theta^2},\quad\hbox{and}\quad \B = \frac{\BB - \kappa\,\theta\,\EE}{1+\kappa^2\,\theta^2}.
\end{equation}
Since this solution in Eq.~(\ref{solution_hatEB}) yields $\EE \cdot \BB = \E \cdot \B = 0$, the electric and magnetic fields propagate orthogonally when in the absence of external sources also in the axion electrodynamics. However, the solution for the electric and magnetic fields is no longer expressed in terms of plane waves, because of the presence of the axion field $\theta$ in Eq.~(\ref{solution_EB}). In fact, the equation for the axion field reduces to the usual Klein-Gordon equation,
\begin{equation}\label{solution_axion}
\left(\Box + m^2\right)\,\theta = 0,
\end{equation}
a solution of which is a wave of momentum $\pp$, frequency $\Omega^2 = |\pp|^2 + m^2$, and amplitude $\theta_0$,
\begin{equation} \label{solution_axion1}
\theta = \theta_0\,\cos(\pp \cdot \rr - \Omega t).
\end{equation}
Notice that, because of the orthogonality between the electric and magnetic fields, the equation of motion for the axion field describes the propagation of a free scalar massive particle along the direction $\pp$. 

\section{Duality between the axion and the electromagnetic field}

\subsection{Gauge transformations in the axion electrodynamics}\label{Gauge transformations in the axion electrodynamics}

In writing the set of Eqs.~(\ref{maxwell_axion_complete_1}), we have expressed the electric and magnetic fields in terms of the vector potential $\A$ and the scalar potential $\phi$ as in Eq.~(\ref{gauge_classic}),
\begin{equation} \label{gauge_classic1}
\E = -\partial_t\,\A - \n\phi,\quad\hbox{and}\quad \B = \n \times {\bf A}.
\end{equation}
However, in the axion electrodynamics theory, these potentials do not directly satisfy the gauge symmetry in the new set of Maxwells' Eq.~(\ref{maxwell_axion1}), and new gauge potentials have to be introduced. To find the gauge field equations in the axion electrodynamics theory, we write the third line in Eq.~(\ref{maxwell_axion1}) as
\begin{equation} \label{gauge_1}
\BB = \B + \kappa \, \theta\,\E = \n \times \Ah,
\end{equation}
where $\Ah$ is a new vector gauge field, and we use the identity $\n \cdot \left(\n \times \Ah \right) = 0$. Using this definition for $\Ah$ in the fourth line in Eq.~(\ref{maxwell_axion1}), and switching the order of the curl and the time derivative, we obtain
\begin{equation}
\n \times \left(\E - \kappa \, \theta\,\B + \partial_t\,\Ah \right) = 0.
\end{equation}
Thanks to the identity $\n \times \n \ph = 0$, valid for any scalar field $\ph$, we obtain
\begin{equation} \label{gauge_2}
\EE = \E - \kappa \, \theta\,\B = -\partial_t\,\Ah - \n \ph.
\end{equation}
Rearranging Eqs.~(\ref{gauge_1}) and~(\ref{gauge_2}), we finally obtain
\begin{equation} \label{EB_gauge}
\begin{array}{l}
(1+\kappa^2\,\theta^2)\,\E = -\partial_t\,\Ah - \n \ph + \kappa\,\theta\,\n\times\,\Ah,\\
(1+\kappa^2\,\theta^2)\,c\B = \n\times\,\Ah + \kappa\,\theta\,\left(\partial_t\,\Ah + \n\ph\right).
\end{array}
\end{equation}
These gauge conditions define the electric and magnetic fields in terms of the gauge potentials $\Ah$ and $\ph$, and reduce to the ordinary electrodynamics potentials in Eqs.~(\ref{gauge_classic1}) when either $\kappa \to 0$ or $\theta \to 0$.

Axion electrodynamics can be reformulated in terms of the gauge potentials. Inserting the gauge Eqs.~(\ref{gauge_1}) and~(\ref{gauge_2}) in the first two lines in Eq.~(\ref{maxwell_axion1}) and rearranging, we obtain the set of wave equations for the axion-electromagnetic field
\begin{equation} \label{axion_em_wave}
\begin{array}{l}
\Box\,\ph = \rho_e + \partial_t\,\left(\n\cdot \Ah + \partial_t \ph\right), \\ 
\Box\,{\Ah} = {\bf J}_e - \n\,\left(\n\cdot \Ah + \partial_t \ph\right), \\ 
\Box\,\theta = -\kappa\,\E\cdot\B - \frac{\partial U(\theta)}{\partial\theta}.
\end{array}
\end{equation}
The first two lines in Eq.~(\ref{axion_em_wave}) resemble the wave equations found in the classical electrodynamics theory, with the only difference lying in the definition of the gauge potentials $\Ah$ and $\ph$ given in Eqs.~(\ref{EB_gauge}). Once the gauge has been fixed so that to fulfill the Lorenz-Lorentz condition
\begin{equation} \label{LLgauge}
\n\cdot \Ah + \partial_t \ph = 0,
\end{equation}
the coupled set of Eqs.~(\ref{axion_em_wave}) in this gauge reads
\begin{equation} \label{axion_em_wave1}
\begin{array}{l}
\Box\,\ph = \rho_e, \\ 
\Box\,\Ah = {\bf J}_e, \\ 
\Box\,\theta = -\kappa\,\E\cdot\B - \frac{\partial U(\theta)}{\partial\theta}.
\end{array}
\end{equation}
Eq.~(\ref{axion_em_wave1}) represents the set of equations for coupled axion-electromagnetic waves in the presence of sources $\rho_e$ and ${\bf J}_e$. 

When switching back to the description of the dynamics in terms of the potentials $\A$ and $\phi$, the analogous of the wave Eqs.~(\ref{axion_em_wave}) are found by arranging Maxwell's source equations in Eq.~(\ref{maxwell_axion1}), giving
\begin{equation} \label{axion_em_wave_1}
\begin{array}{l}
\Box\,\phi = \left[\rho_e +  \kappa\,\n\theta \cdot \B\right] + \partial_t\,\left(\n\cdot \A + \partial_t \phi\right), \\ 
\Box\,\A = \left[{\bf J}_e - \kappa \,(\partial_t\,\theta)\,\B - \kappa\,\n\theta\times\E\right] - \n\,\left(\n\cdot \A + \partial_t \phi\right), \\ 
\Box\,\theta = -\kappa\,\E\cdot\B - \frac{\partial U(\theta)}{\partial\theta}.
\end{array}
\end{equation}
The terms in square brackets in Eq.~(\ref{axion_em_wave_1}) correspond to the new source terms in the presence of the axion field. In fact, Eq.~(\ref{axion_em_wave_1}) can be obtained directly from Eq.~(\ref{axion_em_wave}) by substituting the sources as in Eq.~(\ref{substitution}), while changing $\Ah \to \A$ and $\ph \to \phi$.

\subsection{New conserved currents} \label{Conservation of currents}

We consider the temporal gauge $\phi = 0$, in which $\E = -\partial_t\,\A$, and the last two lines in Eq.~(\ref{axion_em_wave_1}) read
\begin{equation} \label{axion_em_wave2}
\begin{array}{l}
\Box\,\A = {\bf J}_e - \kappa\,\left(\partial_t\,\theta\right)\,\n\times\,\A - \kappa\,\left(\partial_t\,\A\right)\,\times\,\n\theta -\n(\n\cdot\A), \\ 
\Box\,\theta = \kappa\,\B\cdot\,\partial_t\,\A - \frac{\partial U(\theta)}{\partial\theta}.
\end{array}
\end{equation}
We assume that the vector $\A$ is of the form
\begin{equation}\label{choice_A}
\A = A_x\,\ii + B_x\,y\,\kk,
\end{equation}
where $A_x$ depends on time only, while $B_x$ is constant and equal to the magnitude of the magnetic field, $\B = \n\times\A=B_x\,\ii$. 

In this condition, the equations for the $x$-component of $\A$ and for $\theta$ with the axion potential in Eq.~(\ref{quadratic_potential}) read
\begin{equation} \label{axion_em_wave3}
\begin{array}{l}
\Box\,A_x = -\kappa\,B_x\,\dot{\theta} + J_{e,x}, \\ 
\Box\,\theta = \kappa\,B_x\,\dot{A}_x - m^2\,\theta,
\end{array}
\end{equation}
where a dot indicates a partial derivation with respect to time. The equations above show a symmetry between the axion and the electric fields when either $J_{e,x} = -m^2\,A_x$ or when $m = 0$. The first case corresponds to introducing a massive photon term $-m^2\,A^\mu\,A_\mu$ in the Lagrangian, with the photon mass being equal to the axion mass; the second case corresponds to a massless axion. Here, we discuss the first and more general case, which leads to
\begin{equation} \label{axion_em_wave4}
\begin{array}{l}
\left(\Box+m^2\right)\,A_x = -\kappa\,B_x\,\dot{\theta}, \\ 
\left(\Box+m^2\right)\,\theta = \kappa\,B_x\,\dot{A}_x.
\end{array}
\end{equation}
This set of equations is invariant under the symmetry
\begin{equation}\label{O2_symmetry}
\begin{array}{l}
\theta \to \theta + \epsilon\,A_x,\\
A_x \to A_x - \epsilon\,\theta,
\end{array}
\end{equation}
where $\epsilon \ll 1$ is a dimensionless parameter describing the O(2) rotational symmetry between $A_x$ and $\theta$. Following the work by Guendelman~\cite{guendelman}, we write the interaction part of the action as
\begin{equation}
\theta\,\dot{A}_x = \frac{1}{2}\,\left(\theta\,\dot{A}_x - \dot{\theta}\,A_x\right) + \frac{1}{2}\partial_t\left(\theta\,A_x\right),
\end{equation}
so that, once the total derivative term is dropped, the action describing the axion electrodynamics for $\theta$ and the field $A_x$ is
\begin{equation} \label{axion_em_action}
S = \frac{1}{2}\,\int \,\left[ \partial_\mu\theta\,\partial^\mu\theta + \partial_\mu A_x\,\partial^\mu A_x+ \kappa\,B_x\,\left(\theta\,\dot{A}_x - A_x\,\dot{\theta} \right) - m^2\,\left(\theta^2 + A_x^2\right) \right]\,d^4x.
\end{equation}
The Lagrangian term in the equation above is symmetric with respect to the transformation in Eq.~(\ref{O2_symmetry}). Given a generic field $X$ and a Lagrangian $\mathcal{L}$ invariant under an internal symmetry $X \to X + \epsilon \,\delta X$, Noether theorem states that a set of conserved currents is given by
\begin{equation}
J^\mu = -\frac{\partial \mathcal{L}}{\partial\,\left( \partial_\mu X\right)}\,\delta X, \quad\hbox{satisying}\quad \partial_\mu\,J^\mu = 0.
\end{equation}
For the axion electromagnetic action in Eq.~(\ref{axion_em_action}), we obtain
\begin{equation}
\begin{array}{l}
J^0 = \theta\,\dot{A}_x - \dot{\theta}\,A_x + \frac{\kappa\,B_x}{2}\,\left(A_x^2+\theta^2\right),\\
J^i = \theta\,\left(\partial^i\,A_x\right) - \left(\partial^i\theta\right)\,A_x.
\end{array}
\end{equation}
Imposing the conservation of the current, $\partial_\mu J^\mu=0$, gives
\begin{equation}
A_x\,\Box\,\theta -\theta\,\Box\,A_x= \kappa\,B_x\,\left(\theta\,\dot{\theta} + A_x\,\dot{A}_x\right), \\ 
\end{equation}
which can also be obtained by combining the lines in Eq.~(\ref{axion_em_wave4}).

\section{Conclusions}

In Eq.~(\ref{maxwell_axion1}), we have presented a new set of equations describing the evolution of an electromagnetic field in the presence of an axion-like field $\theta$. This set of equations is a generalization of Maxwell's equations when an axion term is present, and it has been derived by requiring its invariance under the dual transformation in Eq.~(\ref{rotation_EM}). This latter request makes this work distinct from previous literature on the subject, in which the dual transformation has not been invoked.

We showed that this set of equations allows for the propagation of a coupled axion-electromagnetic wave, and in Sec.~\ref{Propagation of waves in the axion electrodynamics} we have given a particular solution in which the axion field propagates according to a Klein-Gordon equation, see Eq.~(\ref{solution_axion}), with the orthogonal electric and magnetic fields being expressed in Eq.~(\ref{solution_EB}). Similarly to the classical electrodynamics, the electromagnetic field propagates as a transverse wave; on the contrary, the axion wave behaves as a longitudinal wave.

We have derived the new set of gauge symmetries in the axion electrodynamics, and we have discussed the wave equations in terms of the gauge potentials in Eq.~(\ref{axion_em_wave1}). Using this set of equations, we have derived a new internal symmetry of the axion-electromagnetic field Lagrangian, between the axion field and the gauge potential $\A$, which leads to a conserved current thanks to Noether theorem.

\section*{Acknowledgments}

The author would like to thank Paolo Gondolo (U. of Utah) for useful discussions.


\end{document}